\documentclass[12pt,epsf]{article}
\usepackage{amsmath,amssymb}
\usepackage{bm}
\usepackage{ascmac}
\usepackage{braket}
\usepackage[pdftex]{graphicx}
\graphicspath{{./fig/}}
\usepackage{cite}

\newcommand{\beq}{\begin{equation}}
\newcommand{\eeq}{\end{equation}}
\newcommand{\bea}{\begin{eqnarray}}
\newcommand{\eea}{\end{eqnarray}}

\newcommand{\ba}{\begin{aligned}}
\newcommand{\ea}{\end{aligned}}

\def\pe2{p_E^2}

\begin{document}
\setlength{\baselineskip}{0.7cm}
\begin{titlepage} 
\begin{flushright}
OCU-PHYS 501  \\
NITEP 18
\end{flushright}
\vspace*{10mm}%
\begin{center}{\LARGE\bf Cancellation of One-loop Corrections \\
\vspace*{1mm}
to Scalar Masses in Yang-Mills Theory \\
\vspace*{2mm}
with Flux Compactification}
\end{center}
\vspace*{10mm}
\begin{center}
{\Large Takuya Hirose}$^{a}$ and 
{\Large Nobuhito Maru}$^{a,b}$, 
\end{center}
\vspace*{0.2cm}
\begin{center}
${}^{a}${\it 
Department of Mathematics and Physics, Osaka City University, \\ 
Osaka 558-8585, Japan}
\\
${}^{b}${\it Nambu Yoichiro Institute of Theoretical and Experimental Physics (NITEP), \\
Osaka City University, 
Osaka 558-8585, Japan} 
\end{center}
\vspace*{1cm}

\begin{abstract} 
We calculate one-loop corrections to the mass for the zero mode of scalar field  
 in a six-dimensional Yang-Mills theory compactified on a torus with magnetic flux.
It is shown that these corrections are exactly cancelled 
 thanks to a shift symmetry under the translation in extra spaces. 
This result is expected from the fact 
 that the zero mode of scalar field is a Nambu-Goldstone boson 
 of the translational invariance in extra spaces.  
\end{abstract}
\end{titlepage}

\section{Introduction} 

As one of the approaches to explore the physics beyond the Standard Model (SM), 
 higher dimensional theories have been much attention to paid so far. 
Flux compactification has been extensively studied in string theory \cite{BKLS, IU}. 
Even in the field theories, the flux compactification has been paid much attention. 
Many attractive aspects have been known in such a commpactification: 
 spontaneously supersymmetry breaking \cite{Bachas}, 
 realization of four-dimensional chiral fermion zero-mode without any orbifold \cite{CIM}, 
 explanation of the generation number of the SM fermions \cite{Witten} 
 and computation of Yukawa coupling \cite{CIM, 0903, MS}.


Recently, corrections to the zero mode of the scalar field 
 in effective field theory of flux compactification was studied \cite{B1, Lee, B2}.
According to their surprising results, 
 the corrections at one-loop vanish thanks to the shift symmetry in extra spaces. 
These analyses have been done in an Abelian gauge theory 
 with supersymmetry \cite{B1, Lee} or without supersymmetry \cite{B2}. 

In this paper, 
we extend the work of \cite{B2} to a Non-Abelian gauge theory, that is, 
 a six-dimensional SU(2) Yang-Mills theory compactified on a torus with magnetic flux. 
The cancellation shown in \cite{B2} is nontrivial in this extension 
 since we have to quantize the gauge theory 
 and have to take into account the ghost field contributions, 
 which are irrelevant in an Abelian gauge theory. 
Also, this extension is inevitable to apply to phenomenology. 
We calculate one-loop corrections to the mass 
 for the zero mode of the scalar field from the gauge boson, the scalar field 
 and the ghost field loop contributions. 
These corrections are shown to be exactly cancelled. 
We discuss that a crucial point to show the cancellations is the shift symmetry 
 under translation in extra spaces, as discussed in \cite{B2}.   
We also discuss that the scalar field is a Nambu-Goldstone (NG) boson of the translation in extra spaces. 
This fact implies that the zero mode of the scalar field can only have derivative terms in the Lagrangian 
 and the mass term is forbidden by the shift symmetry.

This paper is organized as follows. 
In section 2, we introduce a six-dimensional Yang-Mills theory with flux compactification.
In section 3, we consider mass eigenvalue and eigenstate for the gauge fields, scalar fields and ghost fields.
An effective Lagrangian in four dimension is derived 
 in section 4.
In section 5, we show the cancellation of corrections to scalar mass at one-loop level.
We discuss the physical reason why the corrections to scalar mass vanish.
We provide our conclusions and discussion in the last section. 
The vertices needed for calculation 
are summarized in Appendix A.


\section{Yang-Mills Theory with Flux Compactification}
We consider a six-dimensional SU(2) Yang-Mills Theory with a constant magnetic flux.
Six-dimensional spacetime is $M^4 \times T^2$, 
 where $M^4$ is a Minkowski spacetime and $T^2$ is a two-dimensional torus.
The Lagrangian of Yang-Mills theory in six dimensions is
\begin{align}
\mathcal{L}_6 &= -\frac{1}{4} F_{MN}^a F^{aMN} \nonumber \\
                      &= -\frac{1}{4} F_{\mu\nu}^a F^{a\mu\nu} - \frac{1}{2} F_{\mu5}^a F^{a\mu5} 
                      -\frac{1}{2} F_{\mu6}^a F^{a\mu6} - \frac{1}{2} F_{56}^a F^{a56},
\label{YMlag}
\end{align}
where the field strength tensor and the covariant derivative are defined by
\begin{align}
F_{MN}^a &= \partial_M A_N^a - \partial_N A_M^a - ig [A_M, A_N]^a, \\
D_M A^a_N &= \partial_M A^a_N + g \varepsilon^{abc} A_M^b A^c_N \nonumber \\
&=\partial_MA^a_N-ig[A_M,A_N]^a.
\end{align}
The spacetime and gauge indices are $M, N = 0,1,\cdots, 6$, $\mu, \nu = 0,1,2,3$, $m, n = 5,6$ and $a = 1,2,3$. 
The metric convention $\eta_{MN} = (-1, +1, \cdots, +1)$ is employed. 
$\varepsilon^{abc}$ is a totally anti-symmetric tensor. 

Let us first discuss how the constant magnetic flux is introduced in our model. 
The magnetic flux is given by the nontrivial background (or vacuum expectation value (VEV)) of 
 the fifth and the sixth component of the gauge field $A_{5,6}$, 
 which must satisfy their classical equation of motion
\bea
D^m \langle F_{mn} \rangle = 0. 
\eea  
In this paper, we choose a solution 
\bea
\langle A_{5}^{1} \rangle = -\frac{1}{2} f x_6,  \qquad 
\langle A_6^1 \rangle = \frac{1}{2} f x_5, \qquad 
\langle A_5^{2,3} \rangle = \langle A_6^{2,3} \rangle = 0,
\eea
which introduces a magnetic field parametrized by a constant $f$, $\langle F^1_{56} \rangle = f$ 
 and breaks a six-dimensional translational invariance spontaneously. 
The magnetic flux is obtained by integrating on $T^2$ and is quantized.
\bea
\frac{g}{2\pi} \int_{T^2} dx_5 dx_6 \langle F^1_{56} \rangle 
 = \frac{g}{2\pi} L^2 f =N \in \mathbb{Z}
\eea
where $L^2$ is an area of the square torus. 
For simplicity, we set $L=1$ hereafter.

It is useful to define $\partial$ and $\phi$ as
\bea
\partial \equiv \partial_z = \partial_5 - i \partial_6, \qquad 
z \equiv \frac{1}{2} (x_5 +i x_6), \qquad \phi = \frac{1}{\sqrt{2}}(A_6 + iA_5).
\label{cc}
\eea
In this complex coordinate, the VEV of $\phi$ is given by $\braket{\phi}=f\bar{z}/\sqrt{2}$, 
 and we expand it around flux background
\bea
\phi^a = \braket{\phi^a} + \varphi^a 
\eea
where $\varphi^a$ is a quantum fluctuation.

The Lagrangian (\ref{YMlag}) can be rewritten by using Eq.~(\ref{cc}) as follows. 
\begin{align}
\mathcal{L}_6 = &-\frac{1}{4} F_{\mu\nu}^a F^{a\mu\nu} 
 - \partial_\mu \bar{\phi}^a \partial^\mu \phi^a 
 -\frac{1}{2} \partial A_\mu^a \bar{\partial} A^{a\mu} 
 + g^2 [A_\mu, \phi]^a [A^\mu, \bar{\phi}]^a \nonumber \\
 & -\frac{i}{\sqrt{2}} (\partial_\mu \phi^a \bar{\partial} A^{a\mu} 
 - \partial_\mu \bar{\phi}^a \partial A^{a\mu}) \nonumber \\
 & +i g \Big\{\partial_\mu \phi^a [A^\mu, \bar{\phi}]^a 
 + \partial^\mu \bar{\phi}^a [A_\mu, \phi]^a \Big\} \nonumber \\
 & -\frac{g}{\sqrt{2}} \Big\{ -\partial A_\mu^a [A^\mu, \bar{\phi}]^a 
 + \bar{\partial} A^{a\mu} [A_\mu, \phi]^a \Big\} \nonumber \\
 &-\frac{1}{4} \Big( D \bar{\phi}^a + \bar{D} \phi^a 
 + \sqrt{2} g [\phi, \bar{\phi}]^a \Big)^2,
\end{align}
where
\begin{align}
D \Phi^a &\equiv (D_5 - iD_6) \Phi^a 
 = \partial \Phi^a - \sqrt{2}g [\phi, \Phi]^a, \\
\bar{D} \Phi^a &\equiv (D_5 + iD_6) \Phi^a 
 = \bar{\partial} \Phi^a + \sqrt{2}g [\bar{\phi}, \Phi]^a,
\end{align}
express the covariant derivatives with respect to the complex coordinates in compactified space. 
$\Phi^a$ denotes an arbitrary field in the adjoint representation.  
We can get rid of the mixing terms between the gauge field and the scalars in the second line of Eq. (\ref{YMlag}) 
 by introducing the gauge-fixing terms, 
\begin{align}
\mathcal{L}_{g-f} = &-\frac{1}{2\xi}(D_\mu A^{a\mu} 
+ \xi \mathcal{D}_m A^{am})^2 \nonumber \\
  = &-\frac{1}{2 \xi} D_\mu A^{a\mu} D_\nu A^{a\nu} 
  - \frac{g}{\sqrt{2}} \partial \bar{\phi}^a [A_\mu, A^\mu]^a 
  +\frac{g}{\sqrt{2}} \bar{\partial} \phi^a [A_\mu, A^\mu]^a 
  + \frac{\xi}{4} (\mathcal{D} \bar{\phi}^a - \bar{\mathcal{D}} \phi^a)^2 \nonumber \\
  & +\frac{i}{\sqrt{2}} (\partial_\mu \phi^a \bar{\partial} A^{a\mu} 
  - \partial_\mu \bar{\phi}^a \partial A^{a\mu}).
\label{gaugefix}
\end{align}
The covariant derivatives $\mathcal{D}, \bar{\mathcal{D}}$ are defined 
 by replacing $\phi^a,\bar{\phi}^a$ in $D, \bar{D}$ 
 with the VEV $\braket{\phi^a}, \braket{\bar{\phi}^a}$, respectively. 

Once we have gauge-fixed, we need to introduce the ghost fields 
 by following Faddeev-Popov procedure to quantize gauge fields.
The ghost Lagrangian reads
\bea
\mathcal{L}_{ghost} =-\bar{c}^a (D_\mu D^\mu + \xi D_m \mathcal{D}^m) c^a.
\label{ghostlag}
\eea
Then, the total Lagrangian is
\begin{align}
\mathcal{L}_{total} = &-\frac{1}{4} F_{\mu\nu}^a F^{a\mu\nu} 
 - \frac{1}{2\xi} D_\mu A^{a\mu} D_\nu A^{a\nu} 
 -\partial_\mu \bar{\phi}^a \partial^\mu \phi^a \nonumber \\
  & -\frac{1}{2} \partial A_\mu^a \bar{\partial} A^{a\mu} 
  + g^2 [A_\mu, \phi]^a [A^\mu, \bar{\phi}]^a 
  -\frac{g}{\sqrt{2}} \Big\{-\partial A_\mu^a [A^\mu, \bar{\phi}]^a 
  + \bar{\partial} A^{a\mu} [A_\mu, \phi]^a \Big\} \nonumber \\
  &+ig \Big\{\partial_\mu \phi^a [A^\mu, \bar{\phi}]^a 
  + \partial^\mu \bar{\phi}^a [A_\mu, \phi]^a \Big\} \nonumber \\
  &-\frac{1}{4} \Big( D \bar{\phi}^a + \bar{D} \phi^a + \sqrt{2}g [\phi, \bar{\phi}]^a \Big)^2 
  + \frac{\xi}{4} (\mathcal{D} \bar{\phi}^a - \bar{\mathcal{D}} \phi^a)^2 \nonumber \\
  &-\bar{c}^a (D_\mu D^\mu + \xi D_m \mathcal{D}^m) c^a.
\label{totallag}
\end{align}

\section{Kaluza-Klein Mass Spectrum}
In this section, we discuss mass eigenstates and eigenvalues of the fields $A_\mu^a, \varphi^a, c^a$, 
 in which it reminds us of a discussion of Landau level in quantum mechanics.

\subsection{Gauge Field}
First, we find mass eigenvalue and eigenstate of the gauge field.
The mass terms of gauge field correspond to the background part of the second line in Eq.(\ref{totallag}).
\begin{align}
\mathcal{L}_{mass}& = -\frac{1}{2} \partial A_\mu^a \bar{\partial} A^{a\mu} 
  + g^2 [A_\mu, \braket{\phi}]^a [A^\mu, \braket{\bar{\phi}}]^a 
  -\frac{g}{\sqrt{2}} \Big\{-\partial A_\mu^a [A^\mu, \braket{\bar{\phi}}]^a 
  + \bar{\partial} A^a_\mu [A^\mu, \braket{\phi}]^a \Big\} \nonumber \\
  & = -\frac{1}{2} \mathcal{D} A_\mu^a \bar{\mathcal{D}} A^{a\mu} \nonumber \\
  & = -\frac{1}{2}A_\mu^a[-\mathcal{D}\bar{\mathcal{D}}]A^{a\mu}.
\end{align}

We note that $\mathcal{D}$ and $\bar{\mathcal{D}}$ can be identified 
 with creation and annihilation operators (see \cite{CIM},\cite{B1}): 
 $a\propto i\bar{\mathcal{D}},a^\dag\propto i\mathcal{D}$.
Expressing them in a matrix form as
\begin{align}
\mathcal{D}^{ac} &=\left(
\begin{array}{ccc}
\partial & 0 & 0 \\
0 & \partial & -\sqrt{2} i \varepsilon^{213} g \braket{\phi^1} \\ 
0 & \sqrt{2} i \varepsilon^{312} g \braket{\phi^1} & \partial \\
\end{array} 
\right)
=\left(
\begin{array}{ccc}
\partial & 0 & 0 \\
0 & \partial & i gf \bar{z}\\ 
0 & -i gf \bar{z} & \partial \\
\end{array} 
\right), \\
\bar{\mathcal{D}}^{ac} &= \left(
\begin{array}{ccc}
\bar{\partial} & 0 & 0 \\
0 & \bar{\partial} & \sqrt{2} i \varepsilon^{213} g \braket{\bar{\phi}^1} \\ 
0 & -\sqrt{2} i \varepsilon^{312} g \braket{\bar{\phi}^1} & \bar{\partial} \\
\end{array} 
\right)
=\left(
\begin{array}{ccc}
\bar{\partial} & 0 & 0 \\
0 & \bar{\partial} & -i gfz \\ 
0 & i gfz & \bar{\partial} \\
\end{array} 
\right), 
\label{matrixDDbar}
\end{align}
we can calculate their commutator
\bea
[i \bar{\mathcal{D}}, i \mathcal{D}]^{ac} 
= \left(
\begin{array}{ccc}
0 & 0 & 0 \\
0 & 0 & -2igf \\ 
0 & 2igf & 0 \\
\end{array} 
\right)
= 2i gf \varepsilon^{a1c}.
\eea
Thus, the creation and annihilation operators can be defined by
\bea
a = \frac{1}{\sqrt{2gf}} i \bar{\mathcal{D}},~~~
a^\dag = \frac{1}{\sqrt{2gf}} i \mathcal{D},
\eea
and we obtain the commutation relation
\bea
[a, a^\dag]^{ac} = i \varepsilon^{a1c}.
\eea
Diagonalizing the covariant derivatives are non-diagonal as
\bea
\mathcal{D}^{ac}_{diag} = \left(
\begin{array}{ccc}
\partial & 0 & 0 \\
0 & \partial - gf \bar{z} & 0 \\ 
0 & 0 & \partial + gf \bar{z} \\
\end{array} 
\right),~~~
\bar{\mathcal{D}}^{ac}_{diag} 
= \left(
\begin{array}{ccc}
\bar{\partial} & 0 & 0 \\
0 & \bar{\partial} + gf z & 0 \\ 
0 & 0 & \bar{\partial} - gf z \\
\end{array} 
\right).
\eea
the commutation relation is diagonalized. 
\bea
[a,a^\dag] 
= 
\left(
\begin{array}{ccc}
0 & 0 & 0 \\
0 & 1 & 0 \\ 
0 & 0 & -1 \\
\end{array} 
\right). 
\eea
Each component of creation and annihilation operators are summarized as follows. 
\bea
\begin{cases}
a_1\equiv\dfrac{1}{\sqrt{2gf}}i\bar{\partial} \\\\
a_2\equiv\dfrac{1}{\sqrt{2gf}}i(\bar{\partial} + gfz) \\\\
a_3\equiv\dfrac{1}{\sqrt{2gf}}i(\bar{\partial} - gfz)
\end{cases}
, \qquad
\begin{cases}
a_1^\dag\equiv\dfrac{1}{\sqrt{2gf}}i\partial \\\\
a_2^\dag\equiv\dfrac{1}{\sqrt{2gf}}i(\partial - gf\bar{z}) \\\\
a_3^\dag\equiv\dfrac{1}{\sqrt{2gf}}i(\partial + gf\bar{z})
\end{cases}.
\label{creation}
\eea
We note that $a_1$ and $a_1^\dag$ play no role of creation and annihilation operators.  
Although $a_2$ and $a_2^\dag$ are ordinary annihilation and creation operators, 
 the roles of creation and annihilation operators for $a_3$ and $a_3^\dag$ are inverted   
 because of $[a_3, a_3^\dag]=-1$. 
The ground state mode functions are determined by $a_2 \psi^2_{0,j}=0, a_3^\dag \psi^3_{0,j}=0$, 
 where 
 $j=0,\cdots,|N|-1$ labels the degeneracy of the ground state.
Higher mode functions $\psi^a_{n_a,j}$ are constructed 
 similar to the harmonic oscillator case, 
\bea
\psi^1_{n_1,j},~~~
\psi^2_{n_2,j} = \frac{1}{\sqrt{n_2!}} (a_2^\dag)^{n_2} \psi^2_{0,j},~~~
\psi^3_{n_3,j} = \frac{1}{\sqrt{n_3!}} (a_3)^{n_3} \psi^3_{0,j},
\eea
and satisfy a orthonormality condition
\bea
\int_{T^2} dx^2 (\psi_{n_a',j'}^{a'})^* \psi_{n_a,j}^a 
 = \delta^{a'a} \delta_{n_a'n_a} \delta_{j'j}
\label{ortho}
\eea
and relations
\begin{align}
\begin{cases}
a_1 \psi^1_{n_1, j} = \psi^1_{n_1, j} \\\\
a_2 \psi^2_{n_2, j} = \sqrt{n_2} \psi^2_{n_2-1, j} \\\\
a_3 \psi^3_{n_3, j} = \sqrt{n_3+1} \psi^3_{n_3+1, j}
\end{cases}
, \qquad
\begin{cases}
a_1^\dag \psi^1_{n_1, j} = \psi^1_{n_1, j} \\\\
a_2^\dag \psi^2_{n_2, j} = \sqrt{n_2+1} \psi^2_{n_2+1, j} \\\\
a_3^\dag \psi^3_{n_3, j} = \sqrt{n_3} \psi^3_{n_3-1, j}
\end{cases}.
\label{creation2}
\end{align}
The mass operator for gauge field (denoted by $\mathcal{H}$ 
 from an analogy of harmonic oscillator Hamiltonian) is diagonalized as 
\bea
\mathcal{H}^{cc'}_{diag} = 
-\mathcal{D}_{diag} \bar{\mathcal{D}}_{diag}
 = 2gf
\left(
\begin{array}{ccc}
n_1 & 0 & 0 \\
0 & n_2 & 0 \\ 
0 & 0 & n_3+1 \\
\end{array} 
\right)
\eea
with Landau level $n_{2,3}$ 
and mass eigenstate of gauge fields are defined by
\bea
\tilde{A}_\mu^a=UA^a_\mu,~~~\tilde{A}^{a\mu}=U^{-1}A^{a\mu}
\label{unitaryrot}
\eea
with a unitary matrix
\bea
U=\frac{1}{\sqrt{2}}
\left(
\begin{array}{ccc}
\sqrt{2} & 0 & 0 \\
0 & 1 & i \\ 
0 & i & 1 \\
\end{array} 
\right).
\eea

\subsection{Scalar Field}
Next, we find mass eigenvalues of the scalar fields.
Extracting quadratic terms for $\varphi^a$ from Eq. (\ref{totallag}), 
 we obtain
\bea
\mathcal{L}_{\varphi\varphi} 
 &=& -\frac{1}{4} \Big(\mathcal{D} \bar{\varphi}^a \mathcal{D} \bar{\varphi}^a 
 + \mathcal{D} \bar{\varphi}^a \bar{\mathcal{D}} \varphi^a 
 + \bar{\mathcal{D}} \varphi^a \mathcal{D} \bar{\varphi}^a 
 + \bar{\mathcal{D}} \varphi^a \bar{\mathcal{D}} \varphi^a 
 - 4gf [\varphi, \bar{\varphi}]^1\Big) \nonumber \\
 &&+ \frac{\xi}{4} 
 \Big(
 \mathcal{D} \bar{\varphi}^a \mathcal{D} \bar{\varphi}^a 
 - \mathcal{D} \bar{\varphi}^a \bar{\mathcal{D}} \varphi^a 
 - \bar{\mathcal{D}} \varphi^a \mathcal{D} \bar{\varphi}^a 
 + \bar{\mathcal{D}} \varphi^a \bar{\mathcal{D}} \varphi^a 
 \Big). 
\label{scalarmass}
 \eea
As the discussion in the previous subsection, we need to diagonalize them. 
In order to justify that the scalar masses can be simultaneously diagonalized 
 by the same unitary rotation 
 \bea
\tilde{\varphi} = U^{-1}\varphi,~~~\bar{\tilde{\varphi}} = U\bar{\varphi}.
\eea 
 as that of the gauge field, 
 we give some arguments below. 
Because of $\mathcal{D} \bar{\varphi}^a \bar{\mathcal{D}} \varphi^a 
 = -\bar{\varphi}^a\mathcal{D}\bar{\mathcal{D}}\varphi^a$, 
 the second and the third terms in the first line of Eq. (\ref{scalarmass}) 
 can be diagonalized by the unitary matrix $U$. 

Next, we focus on the first term in the first line of Eq. (\ref{scalarmass})
\begin{align}
\mathcal{D} \bar{\varphi}^a \mathcal{D} \bar{\varphi}^a 
  &= -\bar{\varphi}^a \mathcal{D} \mathcal{D} \bar{\varphi}^a \nonumber \\
  &= 2gf \Big(\bar{\tilde{\varphi}}^1 (a_1)^2 \bar{\tilde{\varphi}}^1 
  -i \bar{\tilde{\varphi}}^2 (a_2)^2 \bar{\tilde{\varphi}}^3 
  -i \bar{\tilde{\varphi}}^3 (a_3)^2 \bar{\tilde{\varphi}}^2\Big). 
\label{DD}
\end{align}
Integrating out on square torus, the second term and third term in Eq. (\ref{DD}) vanish 
 thanks to the orthogonality of the mode functions.
The first term in Eq. (\ref{DD}) also vanishes 
 since we will consider the zero mode of $\bar{\tilde{\varphi}}^1$ independent of $z, \bar{z}$.
This result is the same for $\bar{\mathcal{D}} \varphi^a \bar{\mathcal{D}} \varphi^a$.
The last term in the first line of Eq. (\ref{scalarmass}) can be also diagonalized 
 by the unitary matrix $U$, 
\bea
-4gf[\varphi,\bar{\varphi}]^1=2\times2gf\bar{\tilde{\varphi}}^a\left(\begin{array}{ccc}
0 & 0 & 0 \\
0 & 1 & 0 \\ 
0 & 0 & -1 \\
\end{array} \right)\tilde{\varphi}^a.
\eea
Applying the same argument to the scalar mass terms from the gauge fixing terms 
 in the second line of Eq. (\ref{scalarmass}), we finally obtain mass eigenvalues of $\varphi^a$:
\bea
m_{\varphi}^2 = gf 
\left(
\begin{array}{ccc}
(1 + \xi) n_1 & 0 & 0 \\
0 & (1 + \xi) n_2+1 & 0 \\ 
0 & 0 & (1 + \xi) n_3 + \xi \\
\end{array} 
\right).
\eea

\subsection{Ghost Field}
Finally, we find mass eigenvalues of ghost field.
Extracting the quadratic terms for $c^a$, we have
\bea
\mathcal{L}_{cc} = -\bar{c}^a \xi \mathcal{D}_m \mathcal{D}^m c^a.
\eea
The differential operator $\mathcal{D}_m\mathcal{D}^m$ can be rewritten as follows. 
\begin{align}
(\mathcal{D}_m \mathcal{D}^m)^{ab} &= (\mathcal{D}_5^2 + \mathcal{D}_6^2)^{ab} \nonumber \\
& = -[(i\mathcal{D}) (i\bar{\mathcal{D}})]^{ab} - \frac{1}{2}[D, \bar{\mathcal{D}}]^{ab} \nonumber \\
& = -2gf \left[ (a^\dag a)^{ab} + \frac{1}{2} i \varepsilon^{a1b} \right],
\end{align}
where we used $[\mathcal{D}_5, \mathcal{D}_6] = [\mathcal{D},\bar{\mathcal{D}}]/2i$.
Thus, the ghost mass matrix is diagonalized as
\bea
m_{c}^{ab} = 2gf
\left(
\begin{array}{ccc}
n_1 & 0 & 0 \\
0 & n_2 + \frac{1}{2} & 0 \\ 
0 & 0 & n_3 + \frac{1}{2} \\
\end{array} 
\right).
\eea
Mass eigenstate of the ghost field is defined as
\bea
\bar{\tilde{c}}^a=U\bar{c}^a,~~~\tilde{c}^a=U^{-1}c^a.
\eea


\section{Effective Lagrangian}
The purpose of this section is to derive effective Lagrangian in four dimensions by KK reduction. 
Although the gauge field $A_\mu^a$ and the ghost field $c^a$ are expanded for all components, 
 only $\varphi^{2,3}$ are expanded as for the scalar field since we are interested in 
 the corrections to the mass for the zero mode of $\varphi^1$ independent of $z, \bar{z}$. 
\begin{align}
\tilde{A}_\mu^a &= \sum_{n_a,j}\tilde{A}_{\mu,n_a,j}^a\psi_{n_a,j}^a~(a=1,2,3), \\
\tilde{\varphi}^{a} &= \sum_{n_a,j} \tilde{\varphi}^a_{n_a,j} \psi^a_{n_a,j},~~~\bar{\tilde{\varphi}}^a=\sum_{n_a,j}\bar{\tilde{\varphi}}^a_{n_a,j}\psi^{a*}_{n_a,j}~(a=2,3) \\
\tilde{c}^a &= \sum_{n_a,j} \tilde{c}^a_{n_a,j} \psi^a_{n_a,j},~~~~\bar{\tilde{c}}^a = \sum_{n_a,j}\bar{c}^a_{n_a,j}\psi^{a*}_{n_a,j}~(a=1,2,3).
\end{align}
Notice that $(A_{\mu,n_a,j}^a)^\dag=A_{\mu,n_a,-j}^a$ is satisfied 
 because the gauge field is real $A_\mu^\dag=A_\mu$ and $(\psi^a_{n_a,j}(z))^*=\psi^a_{n_a,-j}(\bar{z})$ is also satisfied.
From Eq. (\ref{totallag}) and the previous discussion, 
 our Lagrangian is given by\footnote{
 Do not confuse $\tilde{F}_{\mu\nu}$ as the dual of $F_{\mu\nu}$.
 } 
\begin{align}
\mathcal{L}_{total} = &-\frac{1}{4} \tilde{F}_{\mu\nu}^a \tilde{F}^{a\mu\nu} 
  -\partial_\mu \bar{\tilde{\varphi}}^a \partial^\mu \tilde{\varphi}^a 
  - \bar{\tilde{c}}^a \mathcal{D}_\mu \mathcal{D}^\mu \tilde{c}^a \nonumber \\
 & -\frac{1}{2} \tilde{A}^a_\mu \mathcal{H}_{diag} \tilde{A}^{a\mu} 
  -\bar{\tilde{\varphi}}^a m_\varphi^2 \tilde{\varphi}^a - \bar{\tilde{c}}^a m_c^2 \tilde{c}^a \nonumber \\
 &+ ig \Big\{\partial_\mu \varphi^a [A^\mu, \bar{\varphi}]^a + \partial^\mu \bar{\varphi}^a [A_\mu, \varphi]^a \Big\} 
  + g^2 [A_\mu, \varphi]^a [A^\mu, \bar{\varphi}]^a  \nonumber \\
 & - \frac{g}{\sqrt{2}} \Big\{ -\partial A_\mu^a [A^\mu, \bar{\varphi}]^a + \bar{\partial} A^{a\mu} [A_\mu, \varphi]^a \Big\} \nonumber \\
 & + \frac{g}{\sqrt{2}} (\mathcal{D} \bar{\varphi} + \bar{\mathcal{D}} \varphi)^a [\varphi, \bar{\varphi}]^a 
 -\frac{1}{2} g^2 [\varphi, \bar{\varphi}]^a [\varphi, \bar{\varphi}]^a \nonumber \\
 &- \frac{g \xi}{\sqrt{2}} \Big([\varphi, \bar{c}]^a \bar{\partial}c^a - [\bar{\varphi}, \bar{c}]^a \partial c^a \Big). 
 \label{totallag2}
\end{align}
In the above expression, only the quadratic terms in the first and second lines are written in terms of mass eigenstate.
In order to read vertices for Feynman diagram calculations, 
 we must rewrite the remaining interaction terms in terms of the corresponding mass eigenstate, 
 which will be done below.
The relevant vertices required for our calculation in the next section are summarized in Appendix A. 

\subsection{Gauge Field}
In this subsection, the interaction terms including the gauge field are considered. 
It is easy to expand the quartic term. 
\bea
g^2 [A_\mu, \varphi]^a [A^\mu, \bar{\varphi}]^a 
 = -g^2 \varepsilon^{abc} \varepsilon^{ab'c'} 
 \sum_{n_b, j} \sum_{n_b', j'} A_{\mu, n_b, j}^b A_{n_b', j'}^{b' \mu} 
 \varphi^c \bar{\varphi}^{c'} \psi_{n_b, j}^b \psi_{n_b', j'}^{b'},
\eea
then the orthonormality condition for the mode functions leads to
\bea
\mathcal{L}_{\varphi\varphi AA} = -g^2 \varepsilon^{abc} \varepsilon^{ab'c'} \eta^{\mu\nu} 
\sum_{n_b,j} A_{\mu,n_b,j}^b A_{\nu,n_b,-j}^{b'} \varphi^c \bar{\varphi}^{c'}.
\eea
Next, we calculate the cubic term of $\varphi AA$ in a mass eigenstate. 
Expanding $\partial A_\mu^a[A^\mu,\bar{\varphi}]^a$ in components, 
we have 
\begin{align}
\partial A_\mu^a [A^\mu, \bar{\varphi}]^a 
 &=i \varepsilon^{abc} \partial A^a_\mu A^{b\mu} \bar{\varphi}^c \nonumber \\
 &\supset -\partial\tilde{A}^2_\mu \tilde{A}^{2\mu} \bar{\varphi}^1 
  + \partial \tilde{A}^3_\mu \tilde{A}^{3\mu} \bar{\varphi}^1 \nonumber \\
 &= -\left(\frac{2gf}{i}a_2^\dag \tilde{A}^2_\mu 
 + gf \bar{z} \tilde{A}^2_\mu\right) \tilde{A}^{2\mu} \bar{\varphi}^1 
 + \left( \frac{2gf}{i} a_3^\dag \tilde{A}^3_\mu 
 - gf \bar{z} \tilde{A}^3_\mu \right) \tilde{A}^{3\mu} \bar{\varphi}^1
\end{align}
where a symbol $\supset$ in the second line means 
 that only the non-vanishing terms by the orthonormality condition are left. 
In the last line, the partial derivative is replaced by Eq. (\ref{creation}). 
Using the relation (\ref{unitaryrot}) and the orthonormality condition for mode functions, 
 we find
\bea
\mathcal{L}_{\bar{\varphi}AA} 
  = -\sum_{n_2,j} \frac{g\sqrt{\alpha(n_2+1)}}{\sqrt{2}i} 
  \tilde{A}^2_{\mu,n_2,j} \tilde{A}^{2\mu}_{n_2+1,-j} \bar{\varphi}^1 
  +\sum_{n_3,j} \frac{g\sqrt{\alpha(n_3+1)}}{\sqrt{2}i} 
  \tilde{A}^3_{\mu,n_3,j} \tilde{A}^{3\mu}_{n_3+1,-j}\bar{\varphi}^1
\eea
where $\alpha =2gf$.
Similar procedure leads to
\bea
\mathcal{L}_{\varphi AA} = \sum_{n_2,j} 
 \frac{g\sqrt{\alpha(n_2+1)}}{\sqrt{2}i} \tilde{A}^2_{\mu,n_2,j} 
 \tilde{A}^{2\mu}_{n_2+1,-j} \varphi^1 -\sum_{n_3,j} 
  \frac{g\sqrt{\alpha(n_3+1)}}{\sqrt{2}i} \tilde{A}^3_{\mu,n_3,j} 
 \tilde{A}^{3\mu}_{n_3+1,-j}\varphi^1.
\eea
As for the cubic terms $\partial \varphi[A, \bar{\varphi}], \partial \bar{\varphi}[A, \varphi]$, 
 these terms turn out to vanish thanks to the orthgonality condition for mode functions.   
Thus, there is no contribution to the cubic terms in the third line of Eq. (\ref{totallag2}) 
 in four dimensional effective Lagrangian.

\subsection{Scalar Field}
Next, we calculate the cubic and quartic terms for the scalar field.
It is also easy to compute the quartic term. 
\begin{align}
-\frac{1}{2} g^2 [\varphi, \bar{\varphi}]^a [\varphi, \bar{\varphi}]^a 
  &= \frac{1}{2} g^2 \varepsilon^{abc} \varepsilon^{ab'c'} \varphi^b 
  \bar{\varphi}^c \varphi^{b'} \bar{\varphi}^{c'} \nonumber \\
  &= 2 \times \frac{1}{2} g^2 \varepsilon^{abc} \varepsilon^{ab'c'} 
  \sum_{n_b,j} \sum_{n_c,j'} \varphi^b_{n_b,j} \bar{\varphi}^{c'}_{n_{c'},j'} 
  \varphi^{b'} \bar{\varphi}^{c} \psi^b_{n_b,j} \psi^{c'*}_{n_{c'},j'}.
\end{align}
The reason why a facotr 2 appears is that there are two ways to choose a pair of KK expansions: 
 $\varphi^b \bar{\varphi}^{c'}$ or $\varphi^{b'} \bar{\varphi}^c$ 
 since one of the two $\varphi (\bar{\varphi})$ is taken to be $\varphi^1(\bar{\varphi}^1)$. 
Hence, we obtain 
\bea
\mathcal{L}_{\varphi \varphi \varphi \varphi} = g^2 \varepsilon^{abc} \varepsilon^{ab'c'} \delta^{bc'} 
\sum_{n_b,j} \varphi^b_{n_b,j} \bar{\varphi}^{c'}_{n_b,j} \varphi^{b'} \bar{\varphi}^{c}
\eea
The calculation of cubic terms including a single $\varphi^1$ can be done 
 as in the case of gauge field, 
\begin{align}
\mathcal{D} \bar{\varphi}^a [\varphi, \bar{\varphi}]^a 
  &= i \varepsilon^{abc} \mathcal{D} \bar{\varphi}^a \varphi^b \bar{\varphi}^c \nonumber \\
  &\supset \tilde{\varphi}^2 \overline{\tilde{\varphi}}^2 \mathcal{D} \bar{\varphi}^1 
  - \tilde{\varphi}^3 \overline{\tilde{\varphi}}^3 \mathcal{D} \bar{\varphi}^1 
  - \mathcal{D} \overline{\tilde{\varphi}}^2 \tilde{\varphi}^2 \bar{\varphi}^1 
  + \mathcal{D} \overline{\tilde{\varphi}}^3 \tilde{\varphi}^3\bar{\varphi}^1.
\end{align}
The first term and second term vanish because $\mathcal{D} \bar{\varphi}^1=0$ 
 for zero mode of $\varphi^1$. 
Remaining non-vanishing terms are calculated as
\bea
&&\mathcal{L}_{\bar{\varphi} \bar{\varphi} \varphi} 
 = \sum_{n_2, j} \frac{g \sqrt{\alpha(n_2+1)}}{\sqrt{2}i} 
 \overline{\tilde{\varphi}}^2_{n_2+1, j} \tilde{\varphi}^2_{n_2, j} \bar{\varphi}^1 
 - \sum_{n_3, j} \frac{g \sqrt{\alpha(n_3+1)}}{\sqrt{2}i} 
 \overline{\tilde{\varphi}}^3_{n_3, j} \tilde{\varphi}^3_{n_3+1, j} \bar{\varphi}^1, \\
&&
\mathcal{L}_{\bar{\varphi} \varphi\varphi} = - \sum_{n_2, j} \frac{g \sqrt{\alpha(n_2+1)}}{\sqrt{2}i} 
 \tilde{\varphi}^2_{n_2+1, j} \overline{\tilde{\varphi}}^2_{n_2, j} \varphi^1 
 + \sum_{n_3, j} \frac{g \sqrt{\alpha(n_3+1)}}{\sqrt{2}i} \tilde{\varphi}^3_{n_3, j} 
 \overline{\tilde{\varphi}}^3_{n_3+1, j} \varphi^1. 
\eea

\subsection{Ghost Field}
Finally, we compute the cubic terms for the ghost and scalar fields, which include single $\varphi^1$.
\begin{align}
[\varphi, c]^a \bar{\partial} c^a &= i \varepsilon^{abc} 
 \varphi^a \bar{c}^b \bar{\partial} c^c 
\supset 
- \bar{\tilde{c}}^2 \bar{\partial} \tilde{c}^2 \varphi^1 
+ \bar{\tilde{c}}^3 \bar{\partial} \tilde{c}^3 \varphi^1 \nonumber \\
&= -\bar{\tilde{c}}^2 \left( \frac{\sqrt{2gf}}{i} a_2 \tilde{c}^2 
 -gfz \tilde{c}^2 \right) \varphi^1 
 + \bar{\tilde{c}}^3 \left(\frac{\sqrt{2gf}}{i} a_3 \tilde{c}^3 
 + gfz \tilde{c}^3 \right) \varphi^1
\end{align}
where only non-vanishing terms are left in the first line and
 the partial derivative is replaced by Eq.(\ref{creation}). 
Using the relation (\ref{creation2}) and the orthonormality condition for mode functions, 
 we find
\begin{align}
&\mathcal{L}_{\bar{c} c \bar{\varphi}} = \sum_{n_2,j} \frac{g \xi \sqrt{\alpha(n_2+1)}}{\sqrt{2}i} 
 \bar{\tilde{c}}^2_{n_2,j} \tilde{c}^2_{n_2+1,j} \varphi^1 
 - \sum_{n_3,j} \frac{g \xi \sqrt{\alpha(n_3+1)}}{\sqrt{2}i} \bar{\tilde{c}}^3_{n_3+1,j} \tilde{c}^3_{n_3,j} \varphi^1, \\
&\mathcal{L}_{\bar{c} c \varphi} = \sum_{n_2,j} \frac{g \xi \sqrt{\alpha(n_2+1)}}{\sqrt{2}i} 
 \bar{\tilde{c}}^2_{n_2+1,j} \tilde{c}^2_{n_2,j} \bar{\varphi}^1 
 - \sum_{n_3,j} \frac{g \xi \sqrt{\alpha(n_3+1)}}{\sqrt{2}i} \bar{\tilde{c}}^3_{n_3,j} \tilde{c}^3_{n_3+1,j} \bar{\varphi}^1.
\end{align}

\section{Cancellation of One-loop Corrections to Scalar Mass}
In this section, we calculate one-loop corrections to scalar mass for the zero mode of $\varphi^1$ 
 and show that they are exactly cancelled.

\subsection{Gauge Boson Loop}
\begin{figure}[http]
 \begin{center}
  \includegraphics[width=120mm]{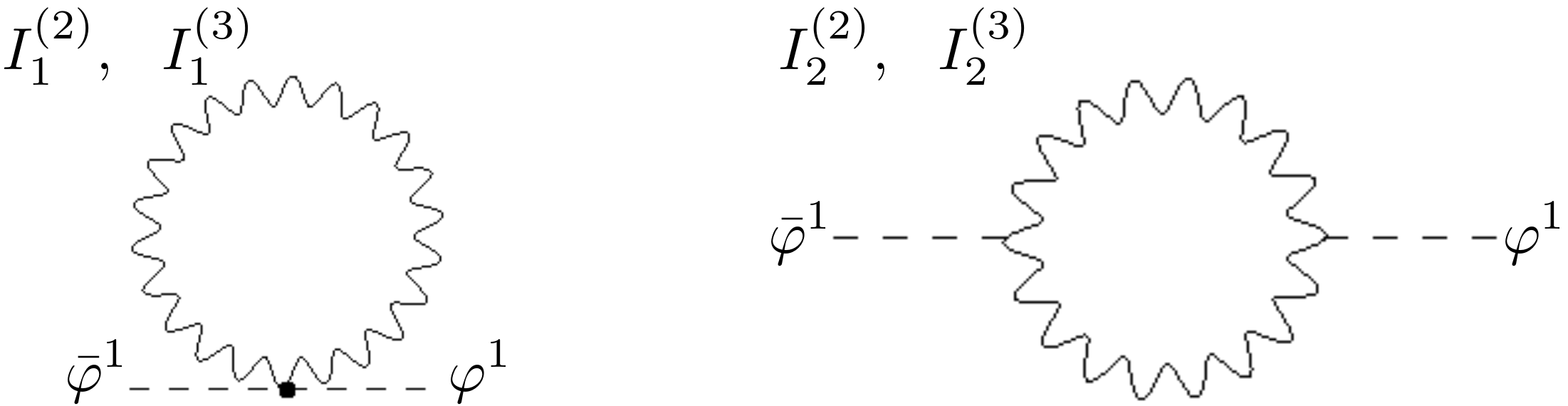}
 \end{center}
 \caption{Gauge boson loop corrections $I_1^{(2,3)}$ and $I_2^{(2,3)}$.}
 \label{one}
\end{figure}

There are two diagrams from the gauge boson loop contributions displayed in Fig. \ref{one}.
Superscript $(2),(3)$ means the contributions from $\tilde{A}_\mu^2, \tilde{A}_\mu^3$ loops, respectively.
Computations of Feynman diagrams are expressed as
\begin{align}
I_1^{(2)} & = -2i g^2 |N| \sum_{n=0}^\infty \int \frac{d^4p}{(2\pi)^4} 
 \left(\frac{3}{p^2 + \alpha n} + \frac{\xi}{p^2 + \alpha n \xi} \right), \\
I_1^{(3)} & = -2i g^2 |N| \sum_{n=0}^\infty \int \frac{d^4p}{(2\pi)^4} 
  \left(\frac{3}{p^2 + \alpha(n+1)} + \frac{\xi}{p^2 + \alpha(n+1) \xi} \right), \\
I^{(2)}_2 & = 2i g^2 |N| \sum_{n=0}^{\infty} \int \frac{d^4p}{(2\pi)^4} 
  \left(\frac{3\alpha(n+1)}{(p^2 + \alpha n)(p^2 + \alpha(n+1))} 
  + \frac{\alpha(n+1) \xi^2}{(p^2 + \alpha n \xi)(p^2 + \alpha(n+1) \xi)} \right), \\
I^{(3)}_2 & =2i g^2 |N| \sum_{n=0}^{\infty} \int \frac{d^4p}{(2\pi)^4} 
  \left(\frac{3\alpha(n+1)}{(p^2 + \alpha(n+1))(p^2 + \alpha(n+2))} \right. \nonumber \\
   & \left. \hspace*{80mm}
   +\frac{\alpha(n+1) \xi^2}{(p^2 + \alpha(n+1) \xi )(p^2 + \alpha(n+2) \xi )} \right), \nonumber \\
\end{align}
where Wick rotation is applied in momentum integrals.\footnote{
In our convention of the metric, the sign of $p^2$ is unchanged after Wick rotation.}
To obtain $I_2^{(2,3)}$, we use a partial fraction decomposition
\bea
\frac{(1 - \xi) p^2}{(p^2 + \alpha n)(p^2 + \alpha n \xi)} = \frac{1}{p^2 + \alpha n} - \frac{\xi}{p^2 + \alpha n \xi}. 
\eea
We now consider the sum of $I_1^{(2)}$ and $I_2^{(2)}$ or $I_1^{(3)}$ and $I_2^{(3)}$. 
\begin{align}
I_1^{(2)} + I_2^{(2)} = &-6i g^2 |N| \sum_{n=0}^\infty \int \frac{d^4p}{(2\pi)^4} 
  \left(\frac{1}{p^2 + \alpha n} - \frac{\alpha(n+1)}{(p^2 + \alpha n)(p^2 + \alpha(n+1))} \right) \nonumber \\
&-2i g^2 |N| \sum_{n=0}^\infty \int \frac{d^4p}{(2\pi)^4} 
  \left(\frac{\xi}{p^2 + \alpha n \xi} - \frac{\alpha(n+1) \xi^2}{(p^2 + \alpha n \xi)(p^2 + \alpha(n+1) \xi)} \right), \\
I_1^{(3)} + I_2^{(3)} = &-6i g^2 |N| \sum_{n=0}^\infty \int \frac{d^4p}{(2\pi)^4} 
  \left(\frac{1}{p^2 + \alpha(n+1)} - \frac{\alpha(n+1)}{(p^2 + \alpha(n+1))(p^2 + \alpha(n+2))} \right) \nonumber \\
&-2i g^2 |N| \sum_{n=0}^\infty \int \frac{d^4p}{(2\pi)^4} 
  \left(\frac{\xi}{p^2 + \alpha(n+1) \xi} - \frac{\alpha(n+1) \xi^2}{(p^2 + \alpha(n+1) \xi)(p^2 + \alpha(n+2) \xi)} \right).
\end{align}
Noting that we can deform the integrand in the first line of $I_1^{(2)}+I_2^{(2)}$ as
\begin{align}
\frac{1}{p^2 + \alpha n} - \frac{\alpha(n+1)}{(p^2 + \alpha n)(p^2 + \alpha(n+1))} 
  &= \frac{1}{p^2 + \alpha n} - (n+1) \left(\frac{1}{p^2 + \alpha n} 
   -\frac{1}{p^2 + \alpha(n+1)}\right) \nonumber \\
  &= -\frac{n}{p^2 + \alpha n} + \frac{n+1}{p^2 + \alpha(n+1)},
\label{subtraction}
\end{align}
we find a crucial result 
\bea
\sum_{n=0}^\infty \int \frac{d^4p}{(2\pi)^4} 
\left(
 -\frac{n}{p^2 + \alpha n} + \frac{n+1}{p^2 + \alpha(n+1)}
\right) 
= 0
\eea
by the shift $n \to n+1$ in the first term. 
The same result holds for the integrand in the first line of $I_1^{(3)} + I_2^{(3)}$. 
As for the integrand in the second line of $I_1^{(2)} + I_2^{(2)}, I_1^{(3)} + I_2^{(3)}$, 
 the same structure can be easily found after the change of variable $p^2 = \xi q^2$. 

Thus, we conclude
\bea
I_1^{(2)} + I_2^{(2)} = 0,~~~I_1^{(3)} + I_2^{(3)} = 0, 
\eea
which implies that the corrections from 
 the gauge boson loop are cancelled.
We emphasize that this cancellation holds for an arbitrary $\xi$.

\subsection{Scalar Loop}
\begin{figure}[http]
 \begin{center}
  \includegraphics[width=120mm]{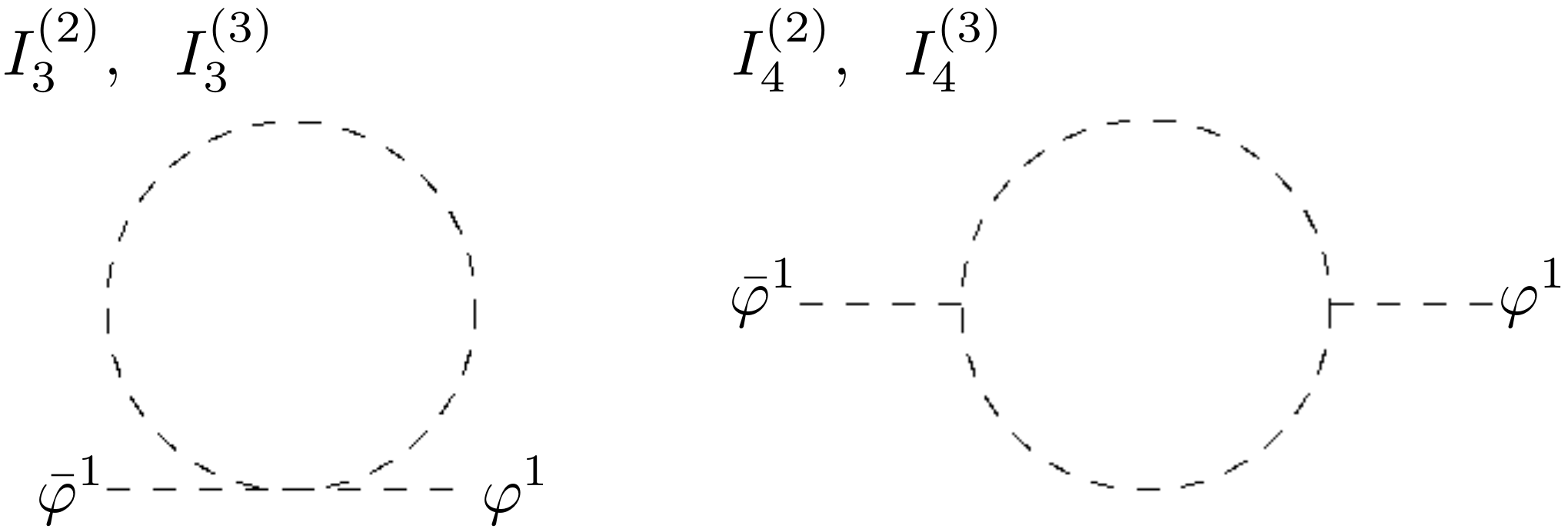}
 \end{center}
 \caption{Scalar loop corrections $I_3^{(2,3)}$ and $I_4^{(2,3)}$.}
 \label{two}
\end{figure}

Two diagrams from the scalar field loop contributions are shown in Fig. \ref{two}.
Here, superscript $(2),(3)$ means the contributions 
 from $\tilde{\varphi}^2, \tilde{\varphi}^3$ loops, respectively.
Computation of Feynman diagrams are expressed as
\begin{align}
I_3^{(2)} &= -i g^2 |N| \sum_{n=0}^\infty \int \frac{d^4p}{(2\pi)^4} 
  \frac{1}{p^2 + \frac{\alpha}{2}((1 + \xi)n +1)}, 
  \label{s1}\\
I_3^{(3)} &= -i g^2 |N| \sum_{n=0}^\infty \int\frac{d^4p}{(2\pi)^4} 
  \frac{1}{p^2 + \frac{\alpha}{2}((1 + \xi)n + \xi)}, \\
I^{(2)}_4 &= \frac{i g^2 |N|}{2} \sum_{n=0}^{\infty} \int\frac{d^4p}{(2\pi)^4} 
  \frac{\alpha(n+1)}{(p^2 + \frac{\alpha}{2}((1 + \xi)n + 1)) (p^2 + \frac{\alpha}{2}((1 + \xi)(n + 1) + 1))}, \\
I^{(3)}_4 &= \frac{i g^2 |N|}{2} \sum_{n=0}^{\infty} \int\frac{d^4p}{(2\pi)^4} 
  \frac{\alpha(n+1)}{(p^2 + \frac{\alpha}{2}((1 + \xi)n +\xi))(p^2 + \frac{\alpha}{2}((1 + \xi)(n + 1) + \xi))},
\label{s4}
\end{align}
where Wick rotation in momentum integral is understood.

\subsection{Ghost Loop}
\begin{figure}[http]
 \begin{center}
  \includegraphics[width=70mm]{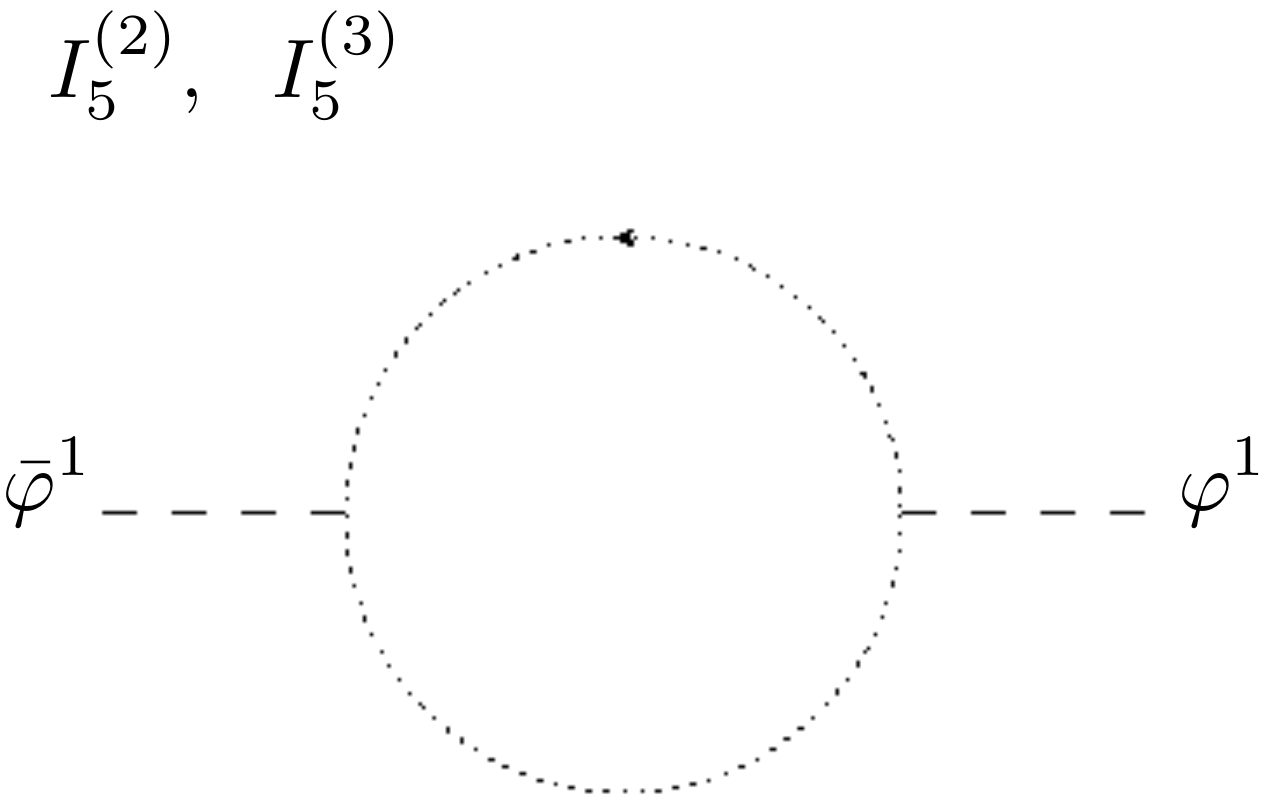}
 \end{center}
 \caption{Ghost loop correction $I_5^{(2,3)}$.}
 \label{three}
\end{figure}

As for the ghost loop contributions, 
 we have only to consider a diagram shown in Fig. \ref{three}.
Superscript $(2),(3)$ means the contributions from $\tilde{c}^2, \tilde{c}^3$ loops, respectively.
Computation of Feynman diagrams are expressed as
\begin{align}
I^{(2)}_5 &= \frac{i g^2 |N| \xi^2}{2} \sum_{n=0}^\infty \int \frac{d^4p}{(2\pi)^4} 
  \frac{\alpha(n+1)}{(p^2 + \alpha (n + \frac{1}{2}))
  (p^2 + \alpha (n + \frac{3}{2}))}, 
  \label{g1}
  \\
I^{(3)}_5 &= \frac{i g^2 |N| \xi^2}{2} \sum_{n=0}^\infty \int\frac{d^4p}{(2\pi)^4} 
  \frac{\alpha(n+1)}{(p^2 + \alpha (n + \frac{1}{2})) 
  (p^2 + \alpha  (n + \frac{3}{2}) )},
\label{g2}
\end{align}
where Wick rotation and a change of variable $p^2 \rightarrow \xi p^2$ are performed in momentum integral.
Notice that we need to consider an overall sign $(-1)$ for the ghost loop.

\subsection{Cancellation between Scalar Loop and Ghost Loop Contributions}
As you can see in subsection 5.1, 
 one-loop corrections to the zero mode scalar mass are cancelled 
 between two diagrams of gauge boson loop.
In this subsection, we show the cancellation between the corrections 
 from the scalar field and the ghost field loops.

First, let us consider the case $\xi=0$.
In this case, the contributions from the ghost field Eq.(\ref{g1}) and (\ref{g2}) trivially vanish 
 since they are proportional to $\xi^2$: $I_5^{(2)}=I_5^{(3)}=0$.
In the $\xi=0$ case, the ghost field is massless, which implies no interaction with scalar fields, 
 therefore it is natural to vanish the contributions from the ghost loops.

Next, let us see how the contributions from the scalar field loop become in $\xi=0$ case.
From Eqs. (\ref{s1}) to (\ref{s4}), the summations from each ghost field contribution can be found
\begin{align}
I_3^{(2)} + I_4^{(2)} &= -i g^2 |N| \sum_{n=0}^\infty \int\frac{d^4p}{(2\pi)^4} 
  \left( \frac{1}{p^2 + \frac{\alpha}{2}(n+1)} 
  - \frac{\frac{\alpha}{2}(n+1)}{(p^2 + \frac{\alpha}{2}(n+1))(p^2 + \frac{\alpha}{2}(n+2))}\right), \\
I_3^{(3)} + I_4^{(3)} &= -i g^2 |N| \sum_{n=0}^\infty \int\frac{d^4p}{(2\pi)^4} 
  \left(\frac{1}{p^2 + \frac{\alpha}{2}n} 
  - \frac{\frac{\alpha}{2}(n+1)}{(p^2 + \frac{\alpha}{2}n)(p^2 + \frac{\alpha}{2}(n+1))}\right).
\end{align}
Utilizing the result Eq. (\ref{subtraction}), 
 we can easily see that these contributions are zero, that is
\bea
I_3^{(2)} + I_4^{(2)}=0,~~~I_3^{(3)}+I_4^{(3)} = 0.
\eea

Let us next consider a more nontrivial case $\xi=1$, 
 in which we expect nontrivial cancellations between 
 the corrections from the scalar field and ghost field loops.
The sum from the scalar and ghost field contributions can be found
 \begin{align}
I_3^{(2)} + I_4^{(2)} + I_5^{(2)} = 
 & -i g^2 |N| \sum_{n=0}^\infty \int \frac{d^4p}{(2\pi)^4} 
  \left( \frac{1}{p^2 + \alpha (n + \frac{1}{2})} \right. \nonumber \\
 & \left. - \frac{1}{2} \frac{\alpha(n+1)}{(p^2 + \alpha \left(n + \frac{1}{2} \right) ) 
 (p^2 + \alpha\left(n + \frac{3}{2} \right) )} 
 - \frac{1}{2} \frac{\alpha(n+1)}{(p^2 + \alpha\left( n + \frac{1}{2} \right) )
 (p^2 + \alpha ( n + \frac{3}{2}) )} \right) \nonumber \\
= & -i g^2 |N| \sum_{n=0}^\infty \int \frac{d^4p}{(2\pi)^4} 
 \left(\frac{1}{p^2 + \alpha(n + \frac{1}{2})} - \frac{\alpha(n+1)}{(p^2 + \alpha \left(n + \frac{1}{2} \right) ) 
 (p^2 + \alpha (n + \frac{3}{2} ) )} \right), \\
I_3^{(3)} + I_4^{(3)} + I_5^{(3)} = & -i g^2 |N| \sum_{n=0}^\infty \int \frac{d^4p}{(2\pi)^4} 
\left(\frac{1}{p^2 + \alpha (n + \frac{1}{2})} \right. \nonumber \\
 & \left. - \frac{1}{2} \frac{\alpha(n+1)}{(p^2 + \alpha (n + \frac{1}{2}) ) 
 (p^2 + \alpha (n + \frac{3}{2}) )} 
 - \frac{1}{2} \frac{\alpha(n+1)}{(p^2 + \alpha (n + \frac{1}{2}) ) 
 (p^2 + \alpha (n + \frac{3}{2}) )} \right) \nonumber \\
 = & -i g^2 |N| \sum_{n=0}^\infty \int \frac{d^4p}{(2\pi)^4} 
 \left(\frac{1}{p^2 + \alpha(n + \frac{1}{2})} - \frac{\alpha(n+1)}{(p^2 + \alpha (n + \frac{1}{2}) ) 
 (p^2 + \alpha (n + \frac{3}{2}) )} \right).
\end{align}
Using the result Eq.(\ref{subtraction}) again, we conclude that these contributions are also zero. 
\bea
I_3^{(2)} + I_4^{(2)} + I_5^{(2)} = 0,~~~I_3^{(3)} + I_4^{(3)} + I_5^{(3)} = 0.
\eea

 \subsection{Physical Reason of Cancellation}
We have shown that one-loop corrections to the zero mode of $\varphi^1$ vanish. 
The physical reason of this remarkable result can be understood 
 from the fact that the zero mode of $\varphi^1$ is a NG boson of translational invariance in extra spaces. 
The transformation of translation in extra spaces are given by
\bea
&&\delta_T A_5^a = (\epsilon_5 \partial_5 + \epsilon_6 \partial_6) 
 \tilde{A}_5^a -\frac{f}{2} \epsilon_6 \delta^{a1}, \\
&&\delta_T A_6^a = (\epsilon_5 \partial_5 + \epsilon_6 \partial_6) 
 \tilde{A}_6^a + \frac{f}{2} \epsilon_5 \delta^{a1}
\eea 
where $\epsilon_{5,6}$ means constant parameters of translations in extra spaces. 
These transformations can be rewritten in complex coordinate
\bea
\delta_T \phi^a = (\epsilon \partial + \bar{\epsilon} \bar{\partial}) \varphi^a 
 + \frac{f}{\sqrt{2}} \bar{\epsilon} \delta^{a1}
 \eea
where $\epsilon \equiv \frac{1}{2}(\epsilon_5 + i\epsilon_6)$. 
Focusing on the zero mode of $\varphi^1$ and 
 noticing $\partial \varphi^1 = \bar{\partial} \varphi^1 = 0$, 
 we find 
\bea
\delta_T \phi^1 = \frac{f}{\sqrt{2}} \bar{\epsilon}
\eea
which is simply reduced to a constant shift symmetry. 
This shows that the zero mode of $\varphi^1$ is a NG boson under the translation in extra spaces. 
Therefore, only the derivative terms of the zero mode of $\varphi^1$ are allowed in the Langrangian 
and it is a natural result that one-loop corrections to the zero mode of $\varphi^1$ vanish. 
It is very interesting to note that 
 the cancellations in the explicit calculations above have been shown 
 by relying on the shift $n \to n+1$, 
 which is a remnant of the shift symmetry discussed in this subsection.

\section{Conclusion and Discussion}
In this paper, we have studied that one-loop corrections 
 to the scalar mass in a six-dimensional Yang-Mills theory compactified 
 on a two-dimensional torus with a constant magnetic flux. 
Having performed KK expansion in terms of mode functions specified by the Landau level 
 for the gauge field, the scaler field originated from the gauge field and the ghost field, 
 the four-dimensional effective Lagrangian was derived.  

Using this effective Lagrangian, we have calculated one-loop corrections to the mass 
 for the zero mode of the scalar field from the gauge boson, the scalar field 
 and the ghost field loop contributions.  
What a remarkable thing is that these corrections are shown to be cancelled. 
As for the gauge boson loop contributions, 
 the cancellation was shown in an arbitrary gauge-fixing parameter $\xi$. 
As for the scalar and the ghost loop contributions, 
 the cancellation is independently realized for each fields in $\xi=0$ case, 
 but the cancellation is not separable and nontrivially established in $\xi=1$ case. 
Crucial point to show the cancellations was the shift of mode index $n \to n+1$ 
 in the momentum integral and mode sum, 
 which is a remnant of the shift symmetry described below.  

The physical reason of cancellation is that the zero mode of the scalar field 
 transforms as a constant shift under the translation in extra spaces. 
Therefore, the scalar field is a NG boson of the translation. 
This fact implies that the zero mode of the scalar field can only have derivative terms in the Lagrangian 
 and the mass term is forbidden by the shift symmetry.


In this paper, we have taken a six-dimensional Yang-Mills theory 
 compactified on a torus with the magnetic flux as an illustration. 
Our results are expected to be true at any order of loop calculations, 
 for any other gauge theories and for any other models in extra even dimensions 
 as far as the zero mode of the scalar field is the NG boson of translation in extra spaces.   
It would be interesting to extend our study along these lines. 

As one of the more interesting phenomenological applications, 
 we hit upon an application to the models of gauge-Higgs unification (GHU)
 where the zero mode of scalar field is identified with the SM SU(2) Higgs doublet. 
Actually, the theory of magnetic flux compactification is, in some sense, 
 an extension of GHU in that the VEV of Higgs boson field 
 depends on extra space coordinates, though it is a constant in GHU. 
As it stands, Higgs boson is massless because of a NG boson. 
We have to introduced some explicit breaking terms of the shift symmetry 
 and then Higgs boson has to be a pseudo NG boson to obtain a mass like a pion. 
If such explicit breaking terms are generated by some electroweak scale dynamics, 
 Higgs boson mass is expected to be an order of the electroweak scale 
 even if the compactification scale is an order of the Planck scale.\footnote
 {
Similar situation is present in a GHU model proposed by one of the authors \cite{LMH}, 
 where a six-dimensional GHU model of a scalar QED compactified on a two-dimensional sphere is considered 
 and one-loop corrections to scalar mass have been shown to vanish. 
This is because the Wilson line loop can be always contractible on a sphere, 
 which means vanishing corrections to mass in a context of GHU.  
 } 
This scenario is very interesting and worth to pursue 
 from the viewpoint that we do not have any evidence of the physics beyond the SM so far. 
Model building along this line would be left for our future work. 
 

\section*{Acknowledgments}
This work is supported in part by JSPS KAKENHI Grant Number JP17K05420 (N.M.).

\appendix
\section{Vertex}
In appendix A, we summarize vertices which needed for calculations in section 4.
\begin{figure}[http]
 \begin{center}
  \includegraphics[width=100mm]{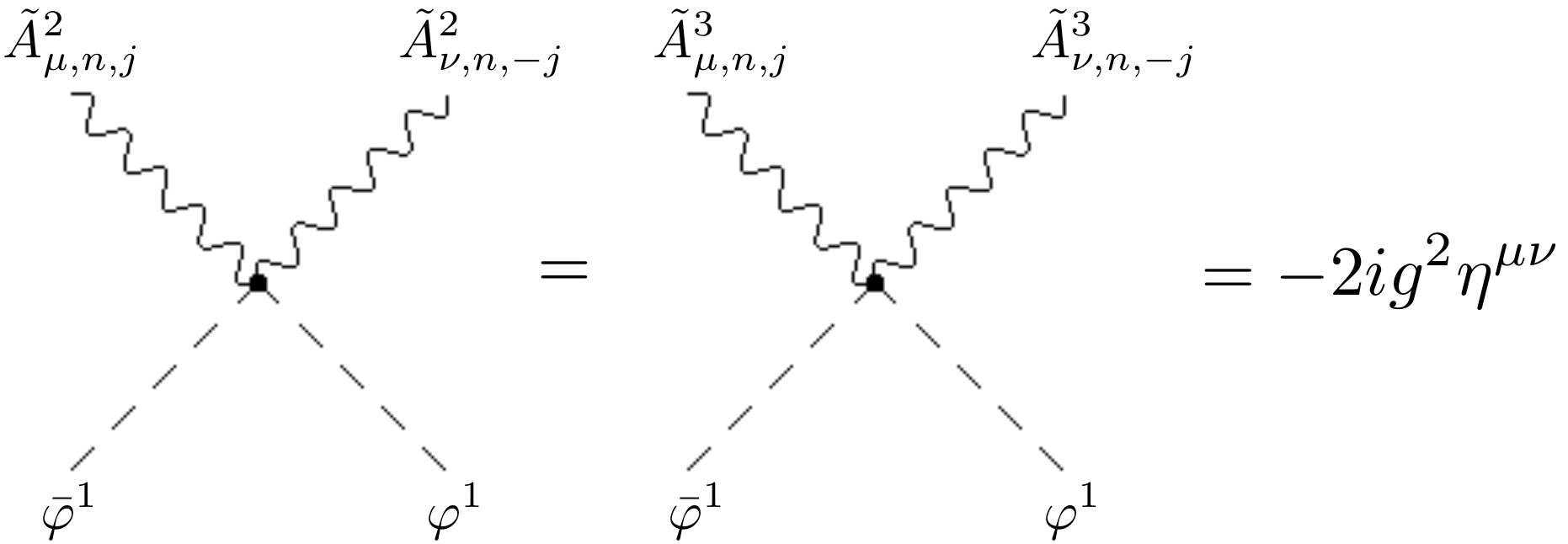}
 \end{center}
 \caption{4-point vertex for $\varphi\varphi AA$}
 \label{SSAA}
\end{figure}
A factor ``2" is a symmetry factor.\footnote{
Note that the minus sign comes form the epsilon tensor. 
}
\begin{figure}[http]
 \begin{center}
  \includegraphics[width=100mm]{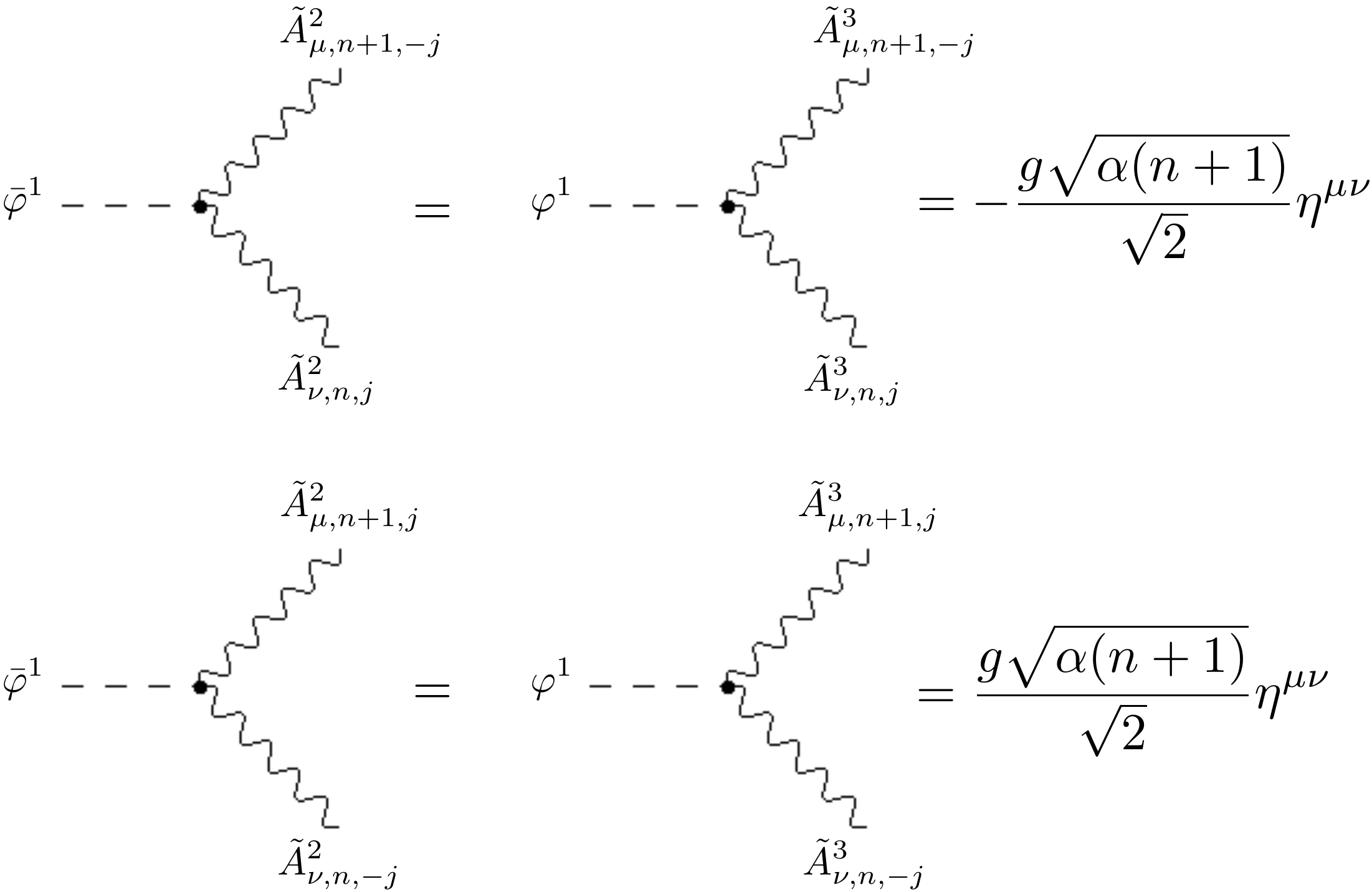}
 \end{center}
 \caption{3-point vertex for $\varphi AA$}
 \label{SAA}
\end{figure}
\begin{figure}[http]
 \begin{center}
  \includegraphics[width=90mm]{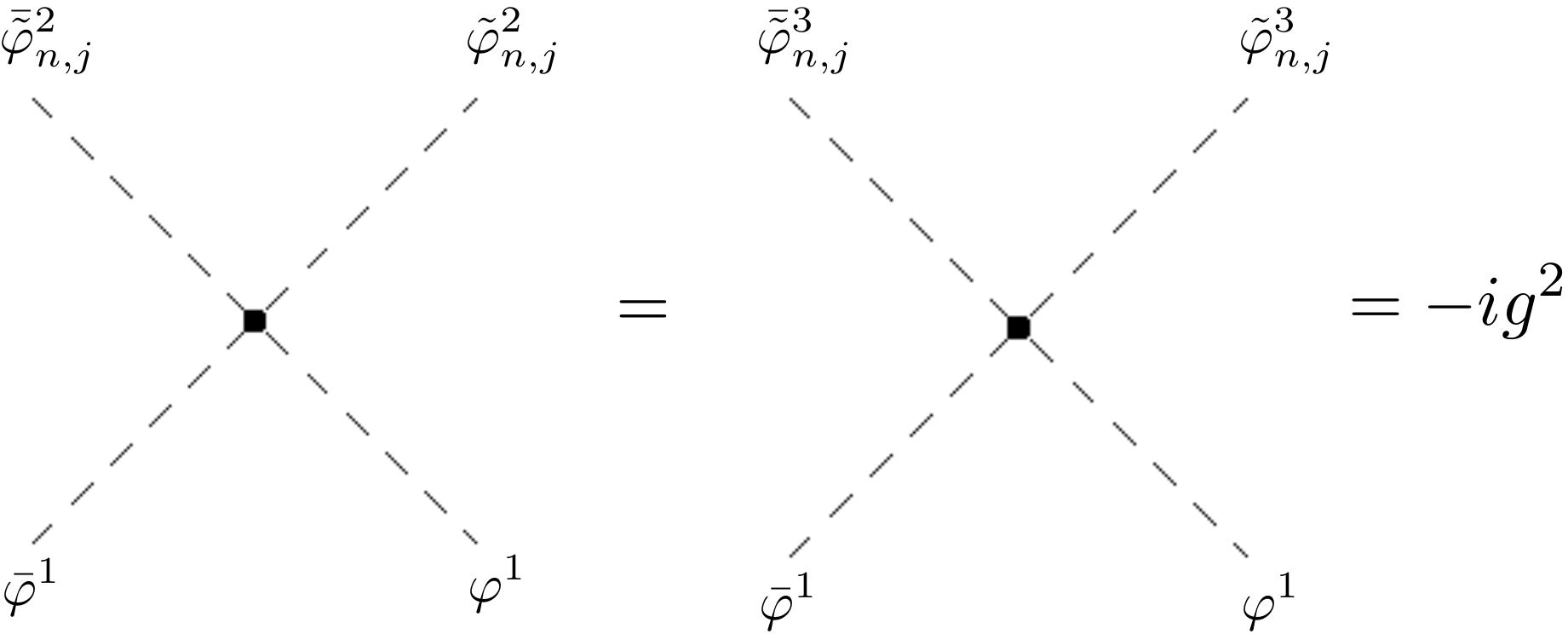}
 \end{center}
 \caption{4-point vertex for $\varphi \varphi \varphi \varphi$}
 \label{SSSS}
\end{figure}
%
%
\begin{figure}[http]
 \begin{center}
  \includegraphics[width=100mm]{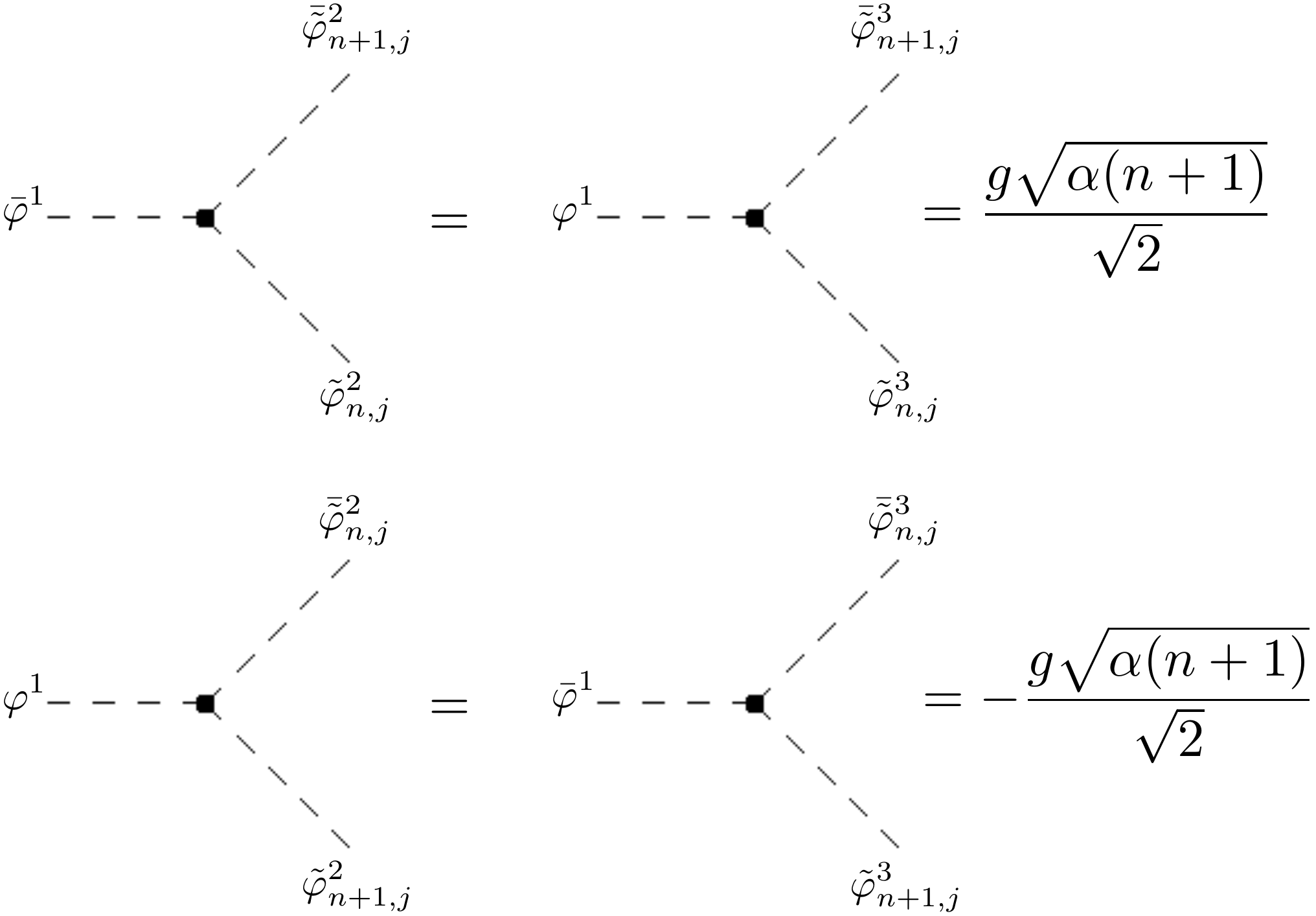}
 \end{center}
 \caption{3-point vertex for $\varphi \varphi \varphi$}
 \label{SSS}
\end{figure}

\begin{figure}[http]
 \begin{center}
  \includegraphics[width=100mm]{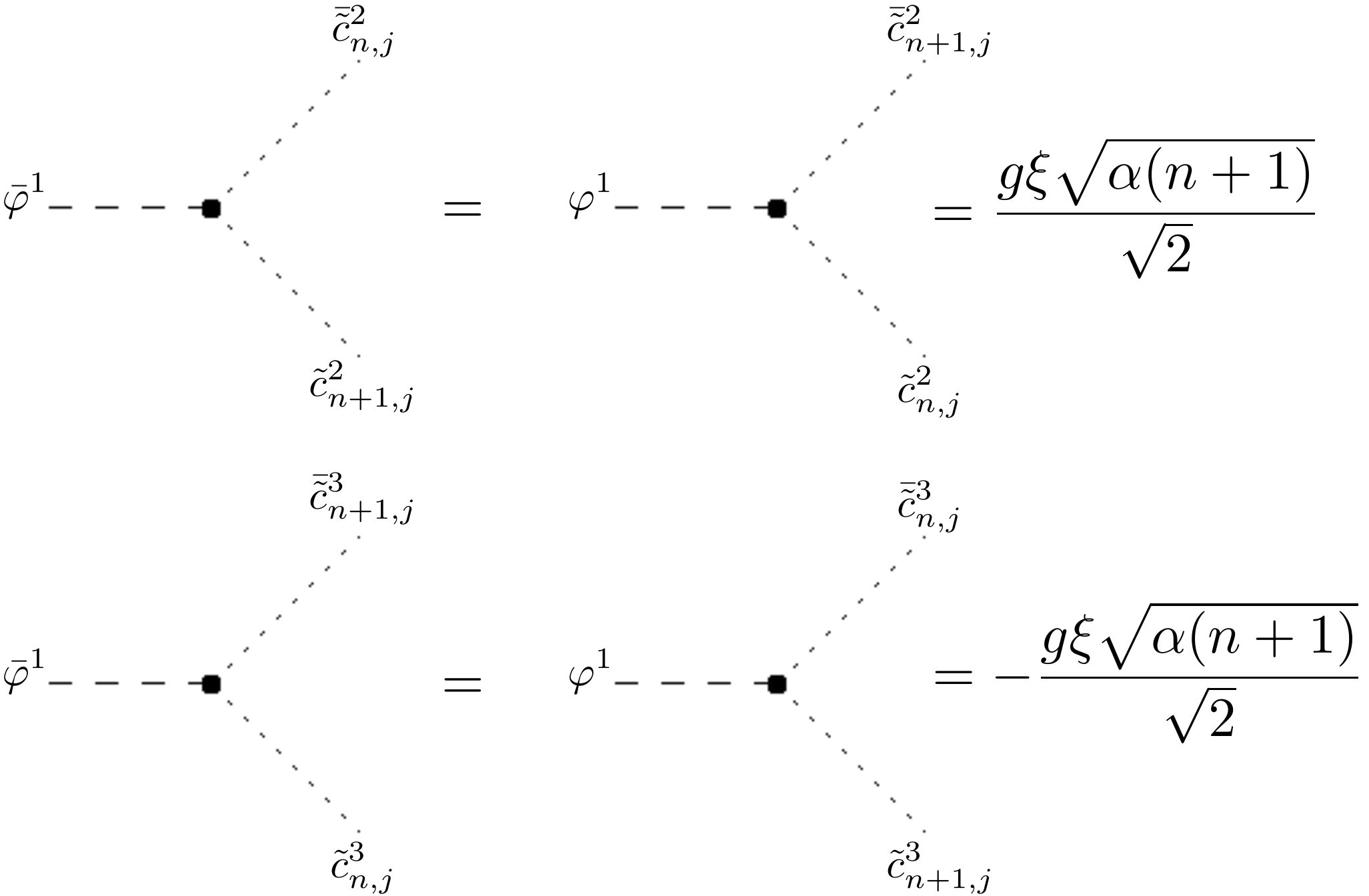}
 \end{center}
 \caption{3-point vertex for $\varphi cc$}
 \label{SCC}
\end{figure}

\newpage


\end{document}